\documentclass[english,twocolumn]{article}

\usepackage{graphicx}
\usepackage{hyperref}
\usepackage{authblk}
\usepackage{float}
\usepackage{bbold}
\usepackage{geometry}
\usepackage[utf8]{inputenc}
\usepackage[backend=biber, style=numeric-comp, sorting=none]{biblatex}
\usepackage{multicol}
\usepackage{amsmath}
\usepackage{bm}
\usepackage{braket}

\usepackage{physics}
\usepackage{color,soul}
\begin{document}

\title{Bandwidth-induced saturation in multimode fiber-based absorbers}

\author{Kfir Sulimany}
\author{Dotan Halevi}
\author{Omri Gat}
\author{Yaron Bromberg}

\affil{Racah Institute of Physics, The Hebrew University of Jerusalem, Jerusalem 91904, Israel}

\date{}
\twocolumn[\begin{@twocolumnfalse}

\maketitle

\begin{abstract}
Multimode fiber-based saturable absorbers enable mode-locking in lasers, generating ultrafast pulses and providing an exceptional platform for investigating nonlinear phenomena. Previous analyses in the continuous wave (CW) limit showed that saturable absorption can be obtained due to nonlinear interactions between transverse modes. We find experimentally that saturable absorption can be achieved thanks to the interplay of single-mode fiber nonlinearity and the wavelength-dependent linear transmission of the multimode fiber, even with negligible intermodal nonlinearities. We further show that even when intermodal nonlinearities are significant, the CW analysis may not be sufficient for long multimode fibers. Understanding the underlying mechanisms of multimode fiber-based saturable absorbers opens new possibilities for developing programmable devices for ultrafast control.
\end{abstract}
\end{@twocolumnfalse}]

A saturable absorber is an optical device with a transmission that depends on the intensity of the incident light, selectively absorbing low-intensity light while allowing transmission of sufficiently high intensities. These devices are extensively used for ultrafast pulse generation in passively mode-locked lasers \cite{grelu2012dissipative, haus2000mode}. When integrated into a laser cavity, high-power pulses stimulated by random fluctuations in the intra-cavity field can be preferentially transmitted by the saturable absorber, while the low-power wings of the pulses are blocked. Over multiple round trips, this process leads to mode-locking.

Traditionally, saturable absorbers are implemented in mode-locked fiber lasers by utilizing semiconductor saturable absorber mirrors \cite{keller1996semiconductor}, as well as carbon nanotubes \cite{set2004laser} and two-dimensional nanomaterials \cite{popa2010sub, sulimany2018bidirectional,klein2018ultrafast}. These methods, however, still possess certain limitations such as low damage thresholds, narrow operational bandwidths, and complex manufacturing processes. Another class of effective saturable absorbers is based on optical nonlinearities, such as nonlinear polarization rotation \cite{matsas1992self}, and nonlinear loop mirrors \cite{doran1988nonlinear, fermann1990nonlinear}. The former method exhibits low environmental stability and its self-starting behavior is challenging, and the latter entails a complex cavity structure. Consequently, a growing research interest is exploring new mechanisms to achieve mode-locking in fiber lasers.

A multimode nonlinear interference-based saturable absorber was theoretically proposed a decade ago by Nazemosadat and Mafi \cite{nazemosadat2013nonlinear}. This device consists of a sandwiched graded-index (GRIN)  multimode fiber spliced at both ends to two single-mode fibers. The device has clear advantages, including low saturation powers, large modulation depths, ultrafast response time, easy fabrication, wide response band, and high damage thresholds \cite{qi2022recent,fu2023recent}. Therefore, in the last decade, dozens of nonlinear interference-based saturable absorbers implemented in ultrafast mode-locked lasers operating at $1.55\mu\mathrm{m}$ \cite{sulimany2022soliton, wang2017er,wang2018stretched,yang2018saturable,chen2018all,zhang2019all,chen2020wavelength,chen2019generation,dong2020er,zhao2019free,li2019band,zhao2018experimental,zhu2019observation,wang2019generation,wu2019femtosecond,gan2021generation,chen2020gimf,chen2019gimf,chen2021generation, zhao2020generation,zhao2020nonlinear,zhang2019c,zhao2018high,zhao2018ultrafast,li2021saturable,lin2021wavelength}, in $1\mu\mathrm{m}$ \cite{teugin2018all,dong2019mode,dong2018mode,pan2021all,dong2018yb,thulasi2020hybrid,dong2019generation,lv2019observation,lv2019nonlinear,gupta2021all,chang2021tunable,chang2021ncf,zhang2020all,chen2021evolution,chang2021generation,thulasi2021all}, and in $2\mu\mathrm{m}$ \cite{li2017mode,li2019continuously,li2019self,jiang2020mode,li2019generation}.

So far, the analysis of multimode fiber-based saturable absorbers has been limited to the continuous wave (CW) limit, where the combination of linear self-imaging in GRIN multimode fibers and nonlinear self-focusing due to the Kerr effect leads to a power-dependent transmission \cite{mafi2011low,nazemosadat2013nonlinear, krupa2019multimode}. This mechanism can drive mode-locking in fiber lasers, yet it neglects spectral effects in multimode fibers and predicts weak saturations for typical peak powers in mode-locked lasers. 

In this work, we study experimentally a multimode fiber-based saturable absorber and analyze its nonlinear features numerically. We find that the interplay between the linear wavelength-dependent transmission of the multimode fiber and the nonlinear spectral broadening in the first single-mode fiber leads to saturable absorption for typical peak powers in mode-locked fiber lasers. Lastly, we study numerically the effect of the nonlinearity in the multimode fiber in our saturable absorber, obtaining insight that paves the way for developing programable saturable absorbers for ultrafast dynamics by all-fiber spectral modulation techniques \cite{finkelstein2023spectral}.

\begin{figure}[ht!]
\begin{centering}
\includegraphics[width=\columnwidth]{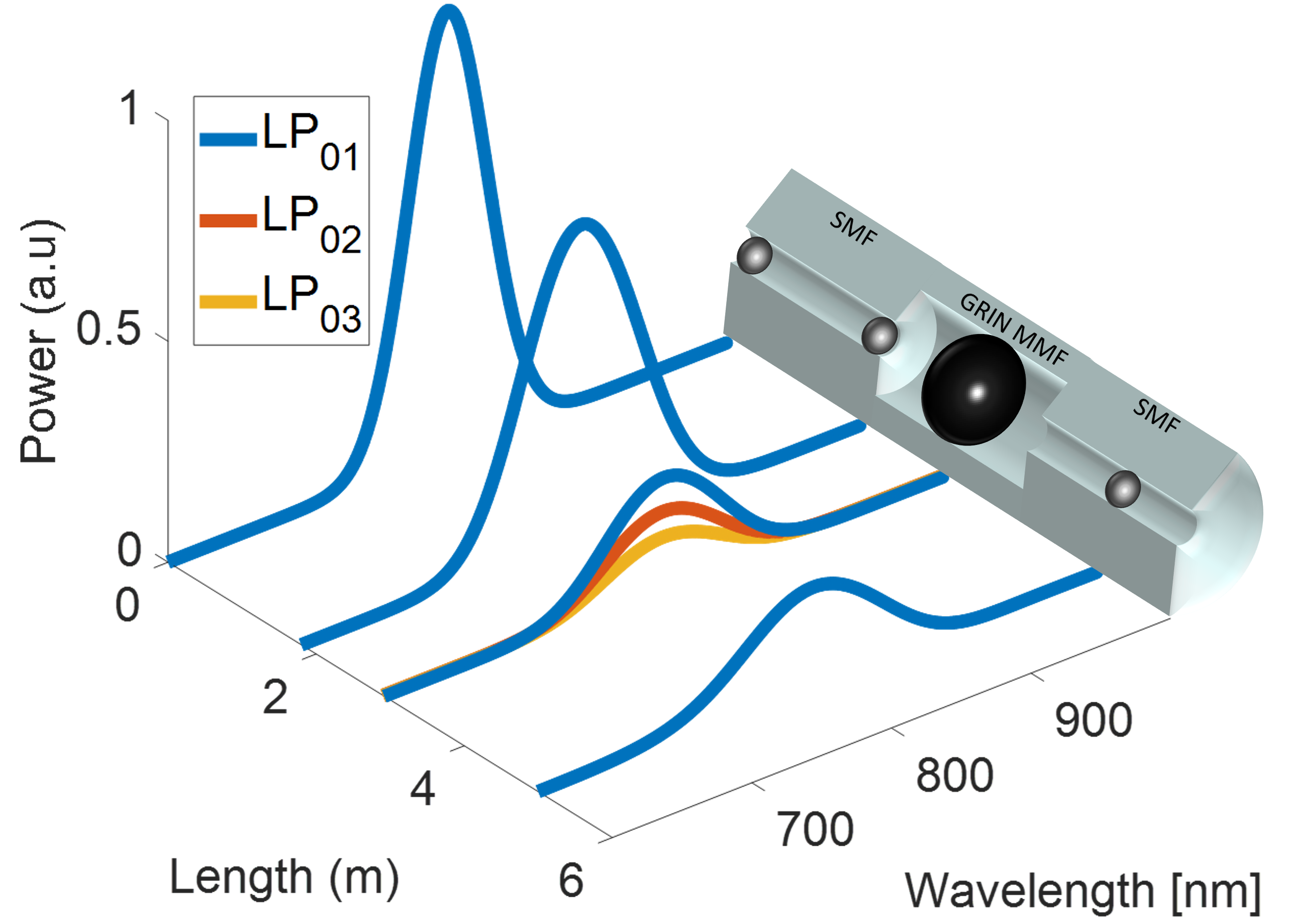}
\par\end{centering}
     \caption{ \textbf{Illustration of pulse propagation in a multimode fiber-based saturable absorber.} An input Gaussian pulse is coupled to a saturable absorber consisting of a graded-index (GRIN) multimode fiber sandwiched between two single-mode fibers (SMF). Self-phase modulation in the first SMF broadens the spectral bandwidth of the input pulse by a few nanometers. When the pulse is coupled to the GRIN multimode fiber, it excites the first few modes of the fiber, labeled by the linearly polarized $LP_{l,m}$ modes. The device transmission, which is determined by the wavelength-dependent coupling to the second SMF, depends on the spectral broadening in the first SMF and is therefore power-dependent. }
 \label{concept}
 \end{figure}

To study the transmission properties of the saturable absorber, we sandwiched a 2\thinspace m -long section of GRIN multimode fiber (OM1 62.5/125 $\mu \mathrm{m}$, Corning) between two 1\thinspace m-long sections of single-mode fiber (SMF, Thorlabs HP780), as illustrated in Fig. \ref{concept}. We inject short pulses from a Ti:Sapphire laser (Coherent Chameleon Ultra II, 680–1060nm, 140fs duration, 80MHz repetition rate) into the device. We illustrate the power spectrum of the pulse in Fig. \ref{concept}. In the first SMF, the input pulse experiences spectral broadening due to self-phase modulation. When the pulse is coupled into the GRIN MMF, a few transverse modes are excited. The coupling to the second SMF now depends on the wavelength-dependent interference between the transverse modes at the output of the GRIN fiber. The transmission through the device thus depends on the spectral broadening in the first SMF and is therefore power-dependent.

\begin{figure}[ht!]
\begin{centering}
\includegraphics[width=\columnwidth]{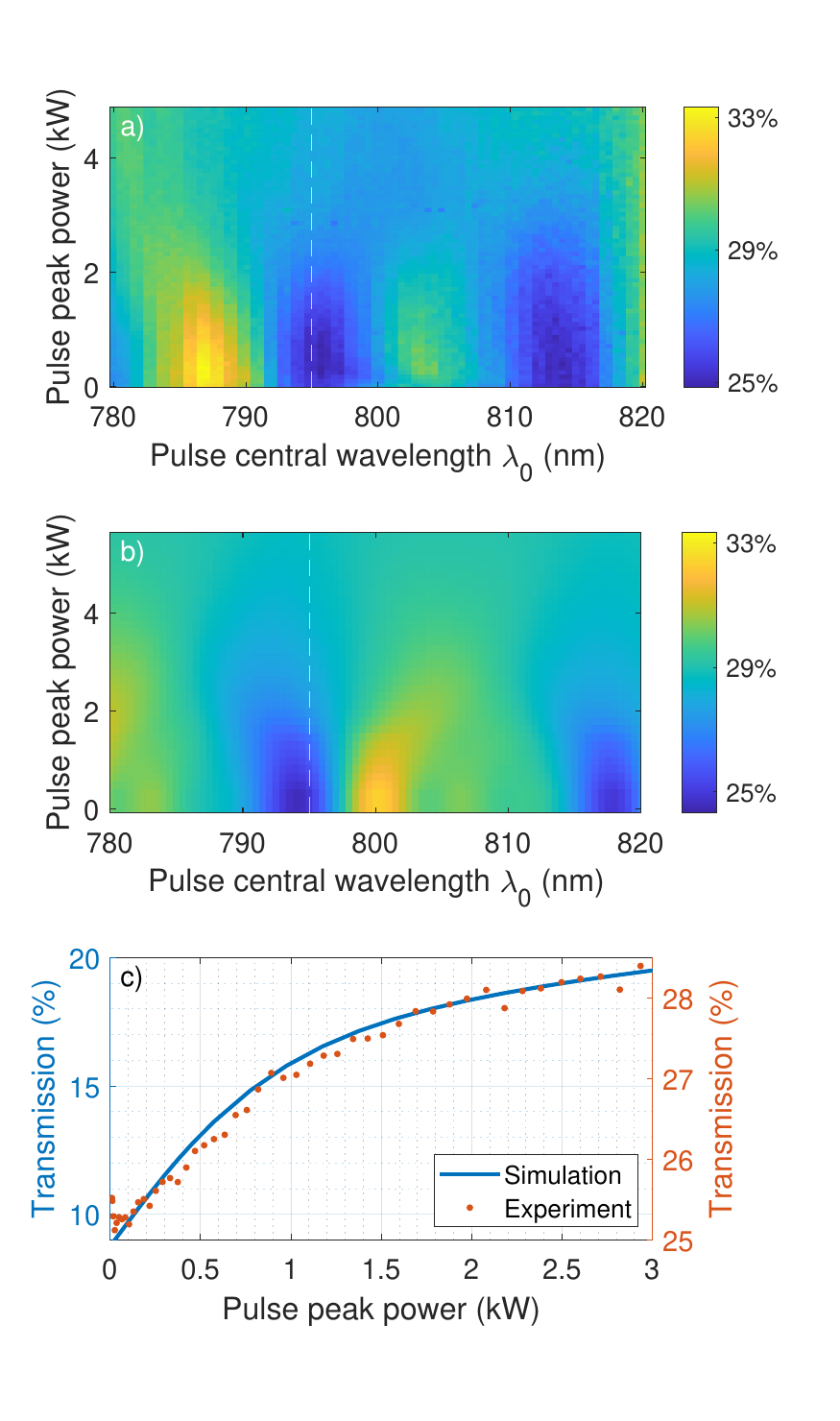}
\par\end{centering}
     \caption{ \textbf{Wavelength-dependent transmission of a multimode fiber-based saturable absorber.} (a) Transmission measurements and (b) simulations as a function of the central wavelength of the pulse $\lambda_0$ and its peak power. At low peak powers, spectral fringes appear due to the phase velocity difference between the modes. These fringes disappear gradually above peak powers of $\sim\mathrm{2kW}$ when the pulse spectrum starts to broaden due to self-phase modulation. Saturated absorption is obtained for central wavelengths that exhibit destructive interference in the linear regime, for example, $\lambda_0=795nm$, marked by the white dashed lines and plotted in (c).}
 \label{EXP}
 \end{figure}

Fig. \ref{EXP}(a) presents the measured power transmitted through the saturable absorber, normalized by input power, as a function of the central wavelength of the pulse $\lambda_0$ and its peak power. At low peak powers, fringes appear in the spectrum due to the phase velocity difference between the modes in the MMF, with a few nanometers scale of change. These fringes disappear above peak powers of $\sim\mathrm{2kW}$, where spectral broadening in the first SMF becomes comparable with the MMF spectral scale of change. The pulse temporal broadening in the SMF, measured by an optical spectrum analyzer (Yokogawa AQ6374), is presented in the supplementary information, section 1. The fading of the fringes leads to saturated absorption in wavelengths that exhibit destructive interference in the linear regime, for example in 795nm marked by the white dashed line. To highlight the saturated absorption, we present in Fig. \ref{EXP}(c) a cross-section of Fig.\ \ref{EXP}(a) at $\lambda_0=795nm$. The saturation power, of less than $1kW$, is a few orders of magnitude lower than that obtained in the continuous wave limit \cite{nazemosadat2013nonlinear}.

To investigate the absorption mechanisms of the SMF-MMF-SMF configuration, we numerically solved the multimode nonlinear Schr\"odinger equation (MM-NLSE) using the numerical solver developed by Wright et al. \cite{wright2017multimode}. The MM-NLSE is given by:
\begin{equation} \label{NLSE}
     \frac{\partial A_{k}}{\partial z} = i\sum_{n} \frac{\beta_n^{(k)}}{n!}(i\frac{\partial}{\partial t})^n A_k + i\frac{n_2\omega_p}{c}\sum_{lmn}S_{klmn}A_lA_mA^*_n
\end{equation}
where $A_k(z,t)$ is the slowly varying amplitude of mode $k$, $z$ is the propagation distance along the fiber, $t$ is time in the comoving frame, and $\beta_n^{(k)} = \partial^n\beta^{(k)}/\partial\omega^n$, $\beta^{(k)}$ being the propagation constant of mode $k$. The nonlinear coupling coefficients $S_{klmn}$ are given by the overlap of the transverse waveforms of the guided modes $F_k(x,y)$: 
\begin{equation}
S_{klmn} = \frac{\int dxdy F_k^*(x,y)F_l^*(x,y)F_m(x,y)F_n(x,y)}{\sqrt{\prod_{i=k,l,m,n} \int dxdy |F_i(x,y)|^2 }}
\end{equation}

To examine the role of nonlinear effects in the multimode fiber, we compare two sets of numerical simulations of Eq. 1. In one set we simulate the full equation and in the other, we neglect its nonlinear terms.  First, we calculated the waveform of a nonlinear $200fs$ pulse propagating in a $1m$ of $4.4 \mu\mathrm{m}$ step-index SMF with an NA of 0.13. Then we calculated the overlap of the field at the output of the SMF with the guided modes of a GRIN multimode fiber with a core diameter $62.5 \mu \mathrm{m}$. For simplifying the numerical computation we considered only the first 11 radial transverse modes of the multimode fiber, to which 97\% of the energy is coupled (see supplementary information section 2 for more details). Next, we calculated the pulse waveform at the end of the multimode fiber, assuming linear propagation as explained above. Lastly, we calculated the coupling of the pulse to another $4.2 \mu \mathrm{m}$ step-index single-mode fiber. 

The calculated pulse transmission as a function of pulse central wavelength and peak power is presented in Fig. \ref{EXP}(b), showing a qualitative agreement with the experimental results. In the experiment, the modulation depth, defined by the difference in the transmission of the high and low peak powers, is lower than in the simulation. The disagreement is due to the dependence of the modulation depth on the multimode fiber length, which was not optimized to fit the experimental results (see supplementary information, section 3). In contrast, above peak powers of $4kW$ the modulation depth becomes insensitive to the pulse peak power, a desired feature for saturable absorbers in mode-locked lasers. Finally, the numerical simulations that include the nonlinear terms in Eq. 1 show that in our configuration, the contribution of nonlinear effects in the multimode fiber is negligible (see supplementary information, section 3).

To expand our investigation to a regime in which intermodal nonlinearity plays an important role, we considered a case similar to the one that has been studied in the CW limit \cite{nazemosadat2013nonlinear}. We initiated a $200 \mathrm{fs}$-wide, $6.9 \mathrm{nJ}$-energy Gaussian pulse, at the start of a GRIN multimode fiber. The occupation of each mode is chosen according to its overlap integral with the SMF mode profile. We considered the first 6 radial transverse modes of the multimode fiber, of which 95\% of the energy is coupled. In this simulation, the pulse was transform-limited at the start of the multimode fiber, so its power was sufficient to induce nonlinearities. We calculated the nonlinear pulse propagation according to Eq. \eqref{NLSE} and then calculated the coupling transmission into a single-mode fiber. 

\begin{figure}[ht!] 
\begin{centering}
\includegraphics[width=1\columnwidth]{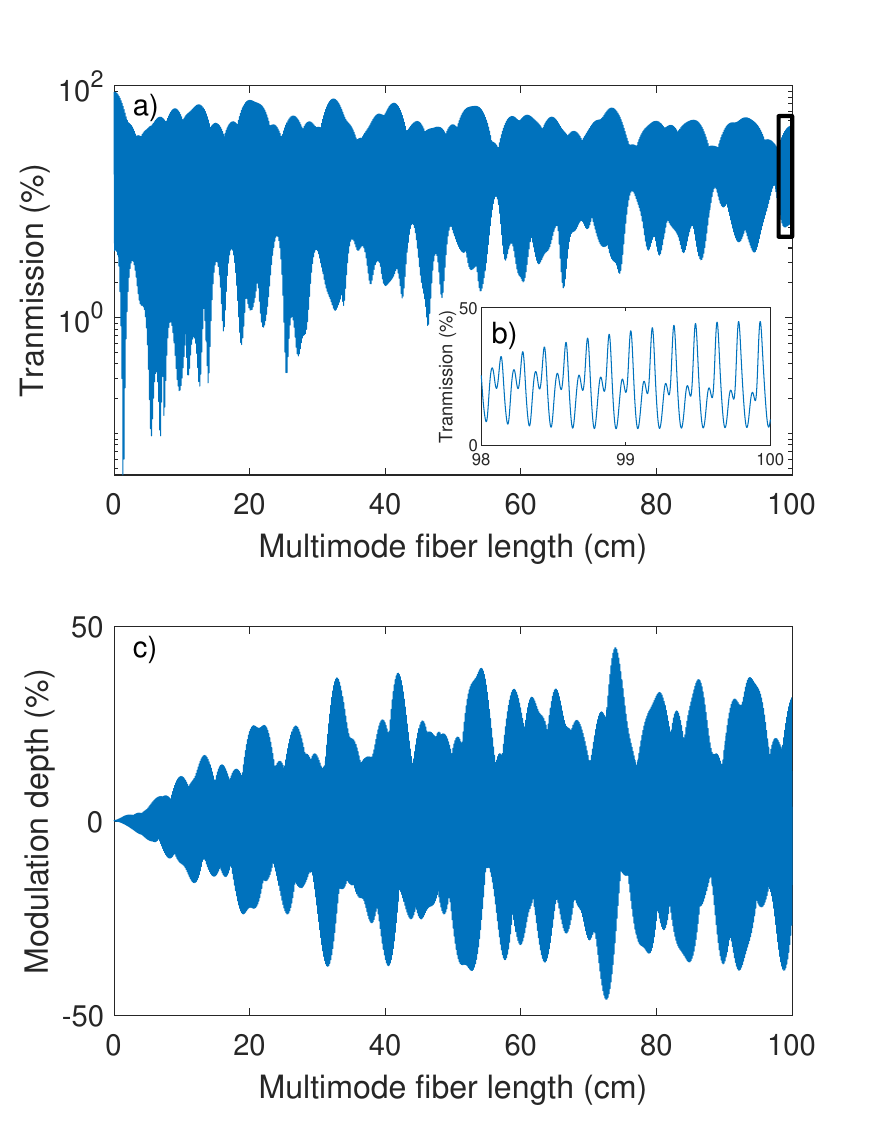}
\par\end{centering}
     \caption{ \textbf{Saturable absorption properties as a function of multimode fiber length.} The calculated coupling transmission of the graded-index (GRIN) multimode fiber to the single-mode fiber, calculated using full Eq. 1, is presented versus the multimode fiber length. As the GIMF multimode fiber length increases, the interplay between its spectral scale of change and the pulse nonlinear spectral broadening prevents the transmission from vanishing. This feature, which is desired in saturable absorbers, cannot be described in the CW limit. The transmission exhibits a fast modulation, as highlighted in the inset, showing $200 \mu \mathrm{m}$ collapse length and $1.5 mm$ revival length. The modulation depth, panel (c) becomes significant after the first few centimeters of the fiber due to intermodal nonlinearities.}
 \label{NUM}
 \end{figure}

The calculated transmission from the GRIN multimode fiber to the SMF versus the GRIN multimode fiber length is presented in Fig.\ \ref{NUM}(a). As the GRIN multimode fiber length increases, the interplay between the spectral scale of change and the pulse nonlinear spectral broadening prevents the transmission from vanishing. This feature, which is desired in saturable absorbers, cannot be captured in the CW limit. The transmission shows fast oscillations, as highlighted in the inset Fig.\ \ref{NUM}(b). In this configuration, the modulation depth depicted in Fig. \ref{NUM}(c), is affected by both the intermodal nonlinear interference and the bandwidth of the multimode fiber. As in the experiment, the modulation depth is calculated by subtracting from the transmission obtained at high peak powers, the transmission obtained at low powers (see supplementary information section 4), and the high power pulse transmission in Fig.\ref{NUM}(a).

Our results regarding the role of spectral effects in multimode fiber-based saturable absorbers suggest that the properties of the saturable absorber are mainly determined by the transmission at weak peak powers, which is mainly determined by linear propagation in the multimode section. It is widely recognized that interference effects in multimode fibers depend sensitively on the multimode fiber length and therefore fine-tuning is required to achieve a desired saturable absorbtion. Recently, we have demonstrated that the linear transmission can be controlled by applying computer-controlled mechanical perturbations \cite{finkelstein2023spectral,shekel2023tutorial}. In the bandwidth-induced multimode fiber-based saturable absorber, therefore, fine-tuning can be achieved by all-fiber modulation. Moreover, in lasers, the optimal desired absorptive nonlinearities for self-starting and steady-state operation are often different, and thus dynamic control allows the implementation of both configurations in a single device.

\section*{Acknowledgments}
The authors wish to thank F. Wise and L. Wright for making freely available the open-source parallel numerical mode solver for the coupled-mode nonlinear Schr\"odinger equations \cite{wright2017multimode}. This research was supported by the \textit{the ISF-NRF Singapore joint research program} (Grant No. 3538/20) and the \textit{Israel Science Foundation (ISF)} (Grant No. 2403/20). K. Sulimany and Y. Bromberg acknowledge the support of the Israeli Council for Higher Education, and the Zuckerman STEM Leadership Program. 

\printbibliography

\clearpage
\section*{Supplementary information}

\section{Spectral broadening in single-mode fibers}
\begin{figure}[ht!]
\begin{centering}
\includegraphics[width=\columnwidth]{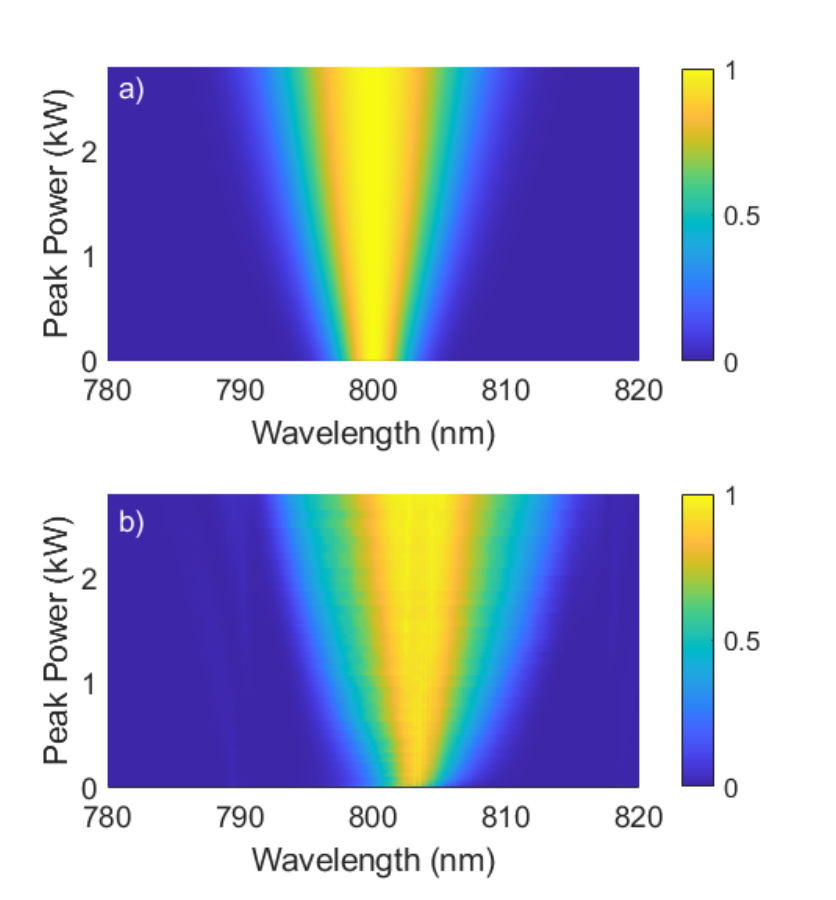}
\par\end{centering}
     \caption{ \textbf{Spectral broadening in a 1m SMF designed for 780nm} a) Normlized spectrum as a function of the pulse peak power, simulated by a pulse propagation modeled by the nonlinear Schrodinger equation. b) Normalized spectrum as a function of pulse peak measured measured by an optical spectrum analyzer (Yokogawa AQ6374).}
 \label{broadning}
 \end{figure}

In order to support our claim that the saturable absorption is the result of spectral broadening in the single-mode fiber (SMF), we present the spectral broadening of a pulse propagating in $1m$ SMF. In Fig. \ref{broadning} (a) we present the simulated normalized spectrum versus the pulse peak power. At roughly 1kW the pulse spectral width is similar to the spectral scale of change in the multimode fiber (MMF). Therefore, at these powers, we get saturation of the absorption. In figure \ref{broadning} (b) we present the normalized spectrum versus the pulse peak power, measured by an optical spectrum analyzer (Yokogawa AQ6374).  

\section{Overlap Integrals}
\begin{figure}[ht!]
\begin{centering}
\includegraphics[width=\columnwidth]{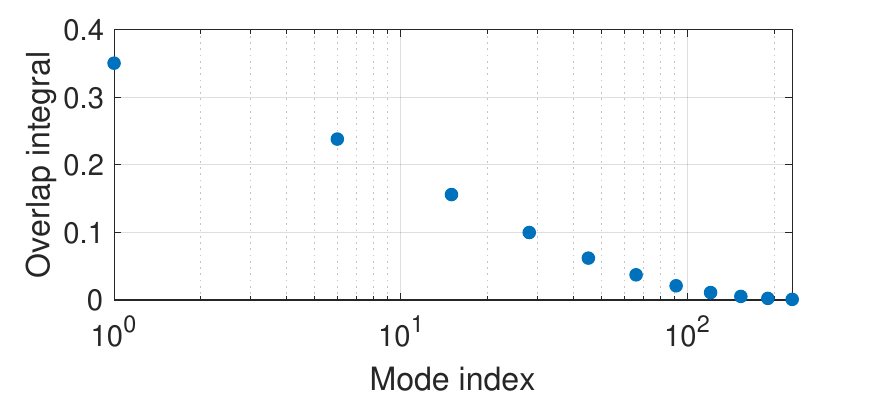}
\par\end{centering}
     \caption{ \textbf{Computed coupling coefficient between the SMF and MMF modes.} The calculated coupling coefficient between the mode of a $4.4\mu \mathrm{m}$ step index SMF with the modes of a $62.5\mu \mathrm{m}$ graded-index MMF. }
 \label{Coupling}
 \end{figure}

To compute the coupling from the SMF to the modes of the MMF, we calculate the overlap integral of the modes:
\begin{equation}
I_{k} = \left(\frac{\int dxdy F_{SMF}^*(x,y)F_k(x,y)}{\sqrt{ \int dxdy |F_{SMF}(x,y)|^2 \int dxdy |F_{k}(x,y)|^2 } }\right)^2
\end{equation}
where $F_{SMF}(x,y)$ is the field of the SMF mode and $F_{k}(x,y)$ is the  $k$th mode of the MMF. In Fig. \ref{Coupling} we present the overlap integrals that have not vanished. 

\section{Bandwidth-induced saturation in multimode fiber-based absorber} \label{Bandwidth-induced saturation in multimode fiber-based absorber}
\begin{figure}[ht!]
\begin{centering}
\includegraphics[width=\columnwidth]{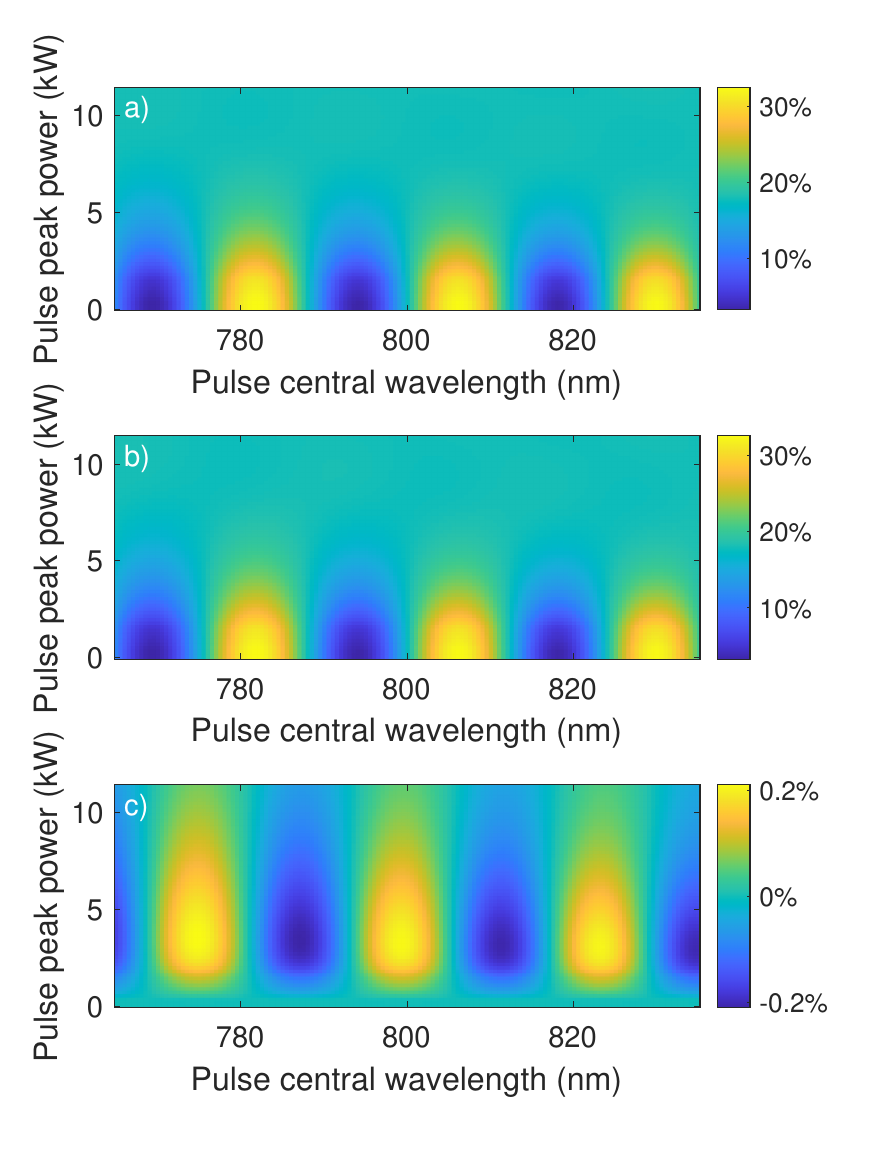}
\par\end{centering}
     \caption{ \textbf{Wavelength-dependent transmission of a multimode fiber-based saturable absorber.} (a) Transmission calculated according to Eq. 1 (b) and for Eq. 1 without the nonlinear terms in the multimode fiber. The difference between the two cases is presented in (c), showing less than $0.2\%$ difference. Therefore the transmission is not dictated by the intermodal nonlinearities, but rather from the MMF bandwidth as explained in the main text.}
 \label{s3}
 \end{figure}

Here we calculate transmission as a function of the pulse peak power and central wavelength, using the coupled nonlinear Schr\"odinger equations (Eq.(1) in the main text). The calculated transmission is presented in Fig. \ref{s3}(a). To study the contribution of nonlinearity in the MMF, we present in Fig. \ref{s3}(b) the transmission computed with the linearized version of Eq.(1). In both cases, we initiated a $200 \mathrm{fs}$-wide Gaussian pulse, at the input of the SMF-MMF-SMF device. Here we considered the first two modes of the MMF. Fig. \ref{s3} (c) presents the difference between panel (a) and panel (b), showing that the saturated absorption is not induced by intermodal nonlinearity in the MMF, but rather from the MMF bandwidth as explained in the main text. 

\begin{figure}[ht!]
\begin{centering}
\includegraphics[width=\columnwidth]{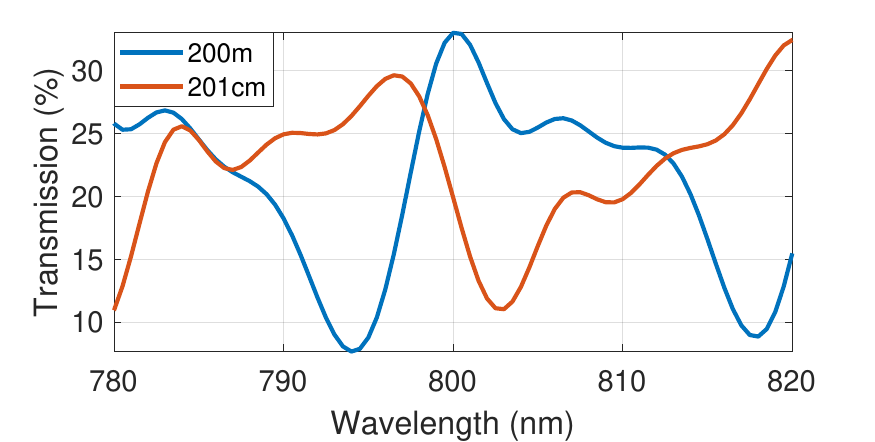}
\par\end{centering}
     \caption{ \textbf{Linear transmission as a function of the central wavelength of the pulse.} The transmission is calculated for low power pulse and presented as a function of the pulse central wavelength for propagation in s $200cm$ (a) and $201cm$ (b) long MMF. The difference between these transmission functions highlights the sensitivity to the fiber length.}
 \label{s4}
 \end{figure}

Last, we calculate the transmission as a function of the peak power of the pulse and its central wavelength, for two multimode fiber lengths Fig. \ref{s4}. (a,b), showing that the modulation depth is different for different multimode fiber lengths.

\section{Intermodal interference and bandwidth-induced saturation in multimode fiber-based absorber}
\begin{figure}[ht!]
\begin{centering}
\includegraphics[width=1\columnwidth]{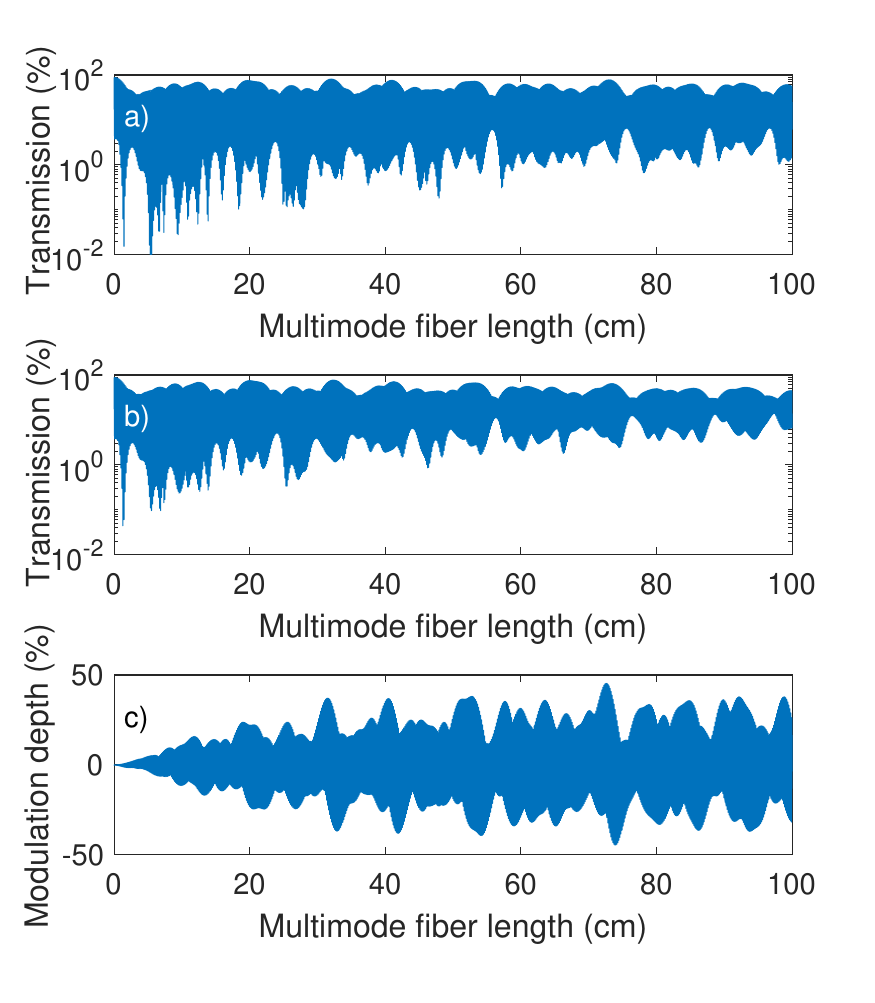}
\par\end{centering}
     \caption{ \textbf{Nonlinear multimode interference-based saturable absorber.} The calculated coupling transmission for a GRIN multimode fiber to the SMF is presented as a function of the multimode fiber length. The transmission is calculated for low peak powers (a), and high peak powers (b). The modulation depth, calculated by the difference between the transmission for high and low peak powers, is presented in (c). In contrast to the configuration presented in the main text, Fig. \ref{s3}, here the saturated absorption is induced by both intermodal nonlinearities and the bandwidth mechanism. }
 \label{s5}
 \end{figure}

Here we calculate transmission as a function of the pulse peak power and its central wavelength $\lambda_0$ using the coupled nonlinear Schr\"odinger equations (Eq. (1) from the main text), for the low peak power regime (Fig. \ref{s5}(a)) and the high peak power regime (Fig. \ref{s5}(b)). In both regimes, we launched a $200 \mathrm{fs}$ pulse, at the input of a GRIN MMF. In panel (c) we plot the difference between the transmissions in (a) and (b). In contrast to the results presented in Fig. 3, here the nonlinear terms in the multimode fiber play an important role in the pulse propagation and the saturated absorption.

\end{document}